\documentclass[pra,aps,twocolumn,amsmath,amssymb]{revtex4}
\usepackage{graphicx}
\pdfoutput=1
\usepackage{amsmath}
\usepackage{color}
\usepackage{float}
\setcitestyle{super}
\usepackage[english]{babel}
\usepackage{color,soul}

\begin{document}

\title{Generation of strain-induced pseudo-magnetic field in a doped type-II Weyl semimetal}

\author{Suman Kamboj$^1$, Partha Sarathi Rana$^2$, Anshu Sirohi$^1$, Aastha Vasdev$^1$, Manasi Mandal$^3$, Sourav Marik$^3$, Ravi Prakash Singh$^3$, Tanmoy Das$^2$}

\author{Goutam Sheet$^1$}
\email{goutam@iisermohali.ac.in}

\affiliation{$^1$Department of Physical Sciences, Indian Institute of Science Education and Research(IISER) Mohali, Sector 81, S. A. S. Nagar, Manauli, PO: 140306, India.}

\affiliation{$^2$Department of Physics, Indian Institute of Science, Bangalore 560012, India}

\affiliation{$^3$Indian Institute of Science Education and Research Bhopal, Bhopal, 462066, India}

\date{\today}

\begin{abstract}

\textbf{In Weyl semimetals, there is an intriguing possibility of realizing a pseudo-magnetic field in presence of small strain due to certain special cases of static deformations. This pseudo-magnetic field can be large enough to form quantized Landau levels and thus become observable in Weyl semimetals. In this paper we experimentally show the emergence of a pseudo-magnetic field ($\sim$ 3 Tesla) by Scanning Tunneling Spectroscopy (STS) on the doped Weyl semimetal Re-MoTe$_2$, where distnict Landau level oscillations in the tunneling conductance are clearly resolved. The crystal lattice is intrinsically strained where large area STM imaging of the surface reveals differently strained domains where atomic scale deformations exist forming topographic ripples with varying periodicity in the real space. The effect of pseudo-magnetic field is clearly resolved in areas under maximum strain.}

\end{abstract}

\maketitle

Weyl semimetals host charge carriers that in many respects behave as massless relativistic particles. In fact, a wide variety of exotic physical phenomena predicted several decades ago in the context of high energy physics have finally been observed in Weyl semimetals.\cite{Weyl1, Hasan, Weyl2, Weyl3, Weyl4, Weyl5, Weyl6, Weyl7} One such example is the so called chiral anomaly which is observed through transport measurements when both electric and magnetic fields are applied to the material.\cite{Hasan_CA, Cava_CA, Park_CA, Pikulin, Zubkov} Realization of Weyl physics in solid state systems has also paved the way for realization of even more exotic phenomena that are rather uncommon in high energy physics. One particular example is the generation of an axial gauge potential $\mathcal{A}$, experimental investigation of which is known to be an extremely complex problem in high energy physics whereas, a synthetic gauge potential is expected to be easily obtained in strained Weyl semimetals.\cite{Pikulin, CKLiu} Such a gauge potential, unlike the usual electromagnetic gauge potential, is an observable as it can give rise to pseudo-electromagnetic fields which may, in principle, interact with fermions of opposite chirality in a Weyl semimetal. It has been theoretically predicted that a pseudo-magnetic field generated by strain in a Weyl semimetal may give rise to the formation of quantized Landau levels and quantum oscillations even in absence of an externally applied magnetic field.\cite{Cortijo_1, Cortijo_2, Franz, Arjona, Ilan, Fujimoto} In fact, one popular school of thought says that Fermi arcs in a Weyl semimetal could be the zeroth Landau levels due to the pseudo-magnetic field generated and localized at the boundary of a crystal. However, a strain-induced pseudo-magnetic field has not been experimentally detected in Weyl semimetals especially because it's strength is often not enough for clear experimental detection. 

In this paper, through scanning tunneling microscopy and spectroscopy down to 300 mK, we show the formation of Landau levels at zero applied magnetic field in intrinsically strained crystals of Re-MoTe$_2$. This is a direct evidence of the generation of a large pseudo-magnetic field in a doped Weyl semimetal. Type-II Weyl fermionic systems are predicted to inherit a set of unique peculiarities in response to a magnetic field due to the tilting of the Weyl dispersions.\cite{TypeIIWeyeLL} Our theoretical analysis reveals that strained Re-doped MoTe$_2$ supports large separation of the Landau-levels (LLs) in the density of states making the experimental detection of the pseudomagnetic field possible.

\begin{figure}[h!]
		\includegraphics[width=0.5\textwidth]{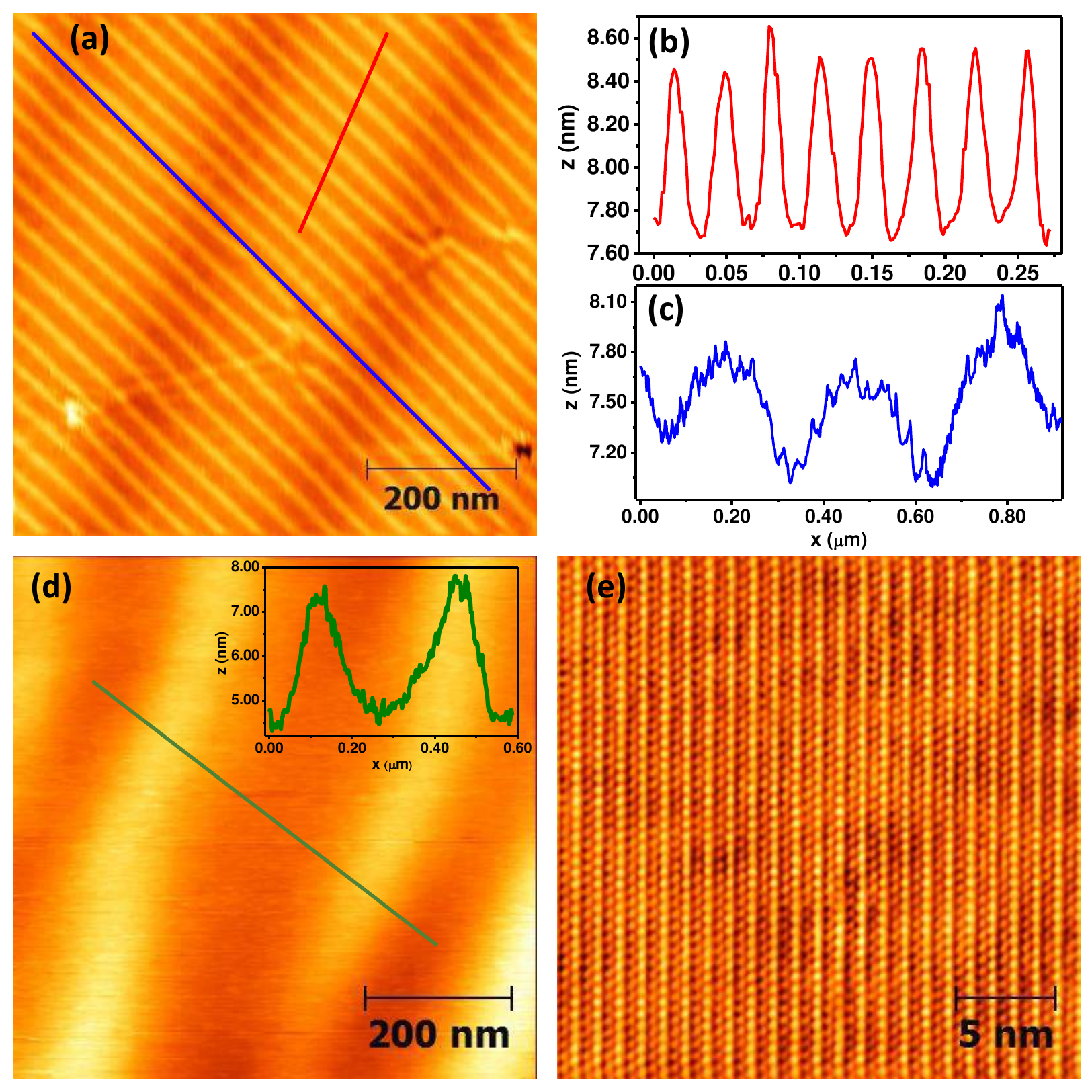}
		\caption{(a) Large area STM topograph of the cleaved surface of Re-MoTe$_2$ obtained in a ``high-strain" region. The periodic modulation due to strain-induced ripples is observed. Variation of topographic height along (b) the red line shown in (a), (c) along the blue line shown in (a). (d) STM topograph of a large area in a ``low-strain" region. The inset shows the topographic modulation along the green line. This modulation due to another bunch of periodic ripples running at an angle with respect to those in (a). (e) STM topograph of a 25 nm x 25 nm area within the same region as in (a). The atoms are clearly resolved.}	

\end{figure}

In the past, following theoretical proposals suggesting the possibility of inducing a large pseudo–magnetic field in graphene,\cite{Graphene} scanning tunneling spectroscopic experiments on highly strained nanobubbles of graphene formed during growth of graphene on a platinum (111) surface revealed Landau level formation corresponding to an extremely high pseudo-magnetic field of the order of 300 Tesla.\cite{Levy} In our single crystals, the strain was possibly induced either during crystal growth or during cleaving and the STM topographs clearly show long range periodic topographic ripples formed due to the strain as it was earlier seen in strained single crystals of Bi$_2$Te$_3$.\cite{Madhavan}
\begin{figure}[h!]
		\includegraphics[width=0.4\textwidth]{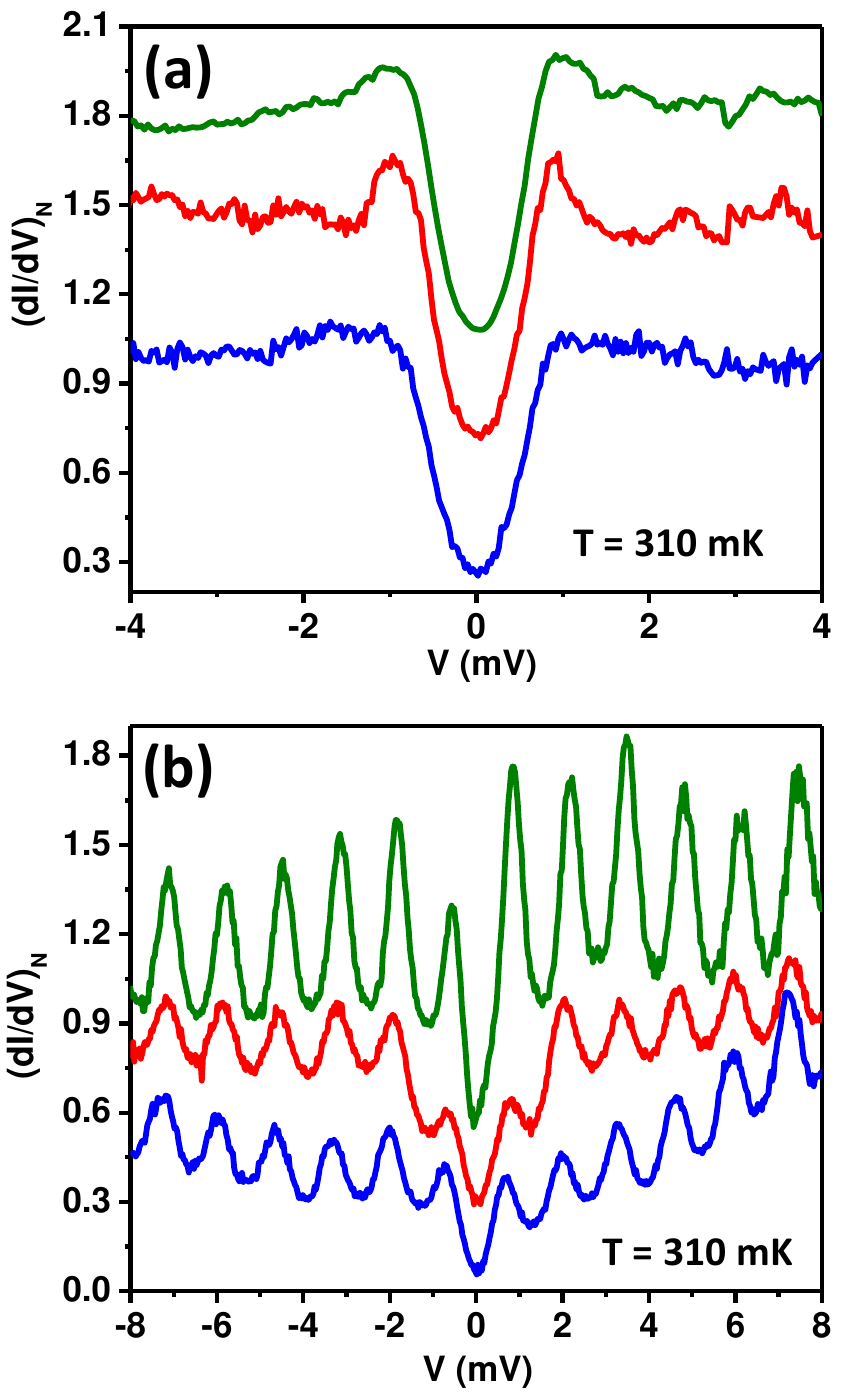}
		\caption{(a) Tunneling spectra recorded at 310 mK at three different points in the ``low-strain" area as shown in Figure 1(d). These show the existence of a superconducting gap. (b) Tunneling spectra in the "high-strain" region at 310 mK showing LL oscillations along with a low bias conductance dip at three different points in the ``high-strain" region as in Figure 1(a).}	

\end{figure}

MoTe$_2$ is an interesting member of the family of the transition metal dichalcogenides because of the multiple structural phases in which the compound can exist.\cite{CFelser, Guguchia} At 250 K, MoTe$_2$ shows a structural phase transition where a high temperature inversion symmetric monoclinic phase finds a new ground state with non-centrosymmetric orthorhombic structure (space group: $Pmn21$, T$_d$  Phase).\cite{Clarke} The Orthorhombic T$_d$ phase is a Weyl semimetal\cite{XDai, Kaminski} which is also known to show superconductivity below a very low critical temperature ($T_c$) of 0.1\,K.\cite{Qi} By introducing strain in the crystals of MoTe$_2$ through large external hydrostatic pressure, it has also been possible to enhance the $T_c$ of MoTe$_2$ significantly. Recently it was shown that the same high $T_c$ phase of MoTe$_2$ can also be achieved by doping the Mo-sites of MoTe$_2$ with Re.\cite{RPSingh} At 20\% Re doping, MoTe$_2$ shows a dramatically increased transition temperature as high as 3.8\,K at ambient pressure.\cite{RPSingh} While the mechanism through which the enhancement of $T_c$ in Re-MoTe$_2$ occurs is not understood as of now, from the similarities of the results with those obtained under high pressure, it is rational to consider chemically induced or otherwise developed internal strain as one of the highly probable mechanisms. And, if the lattice gets strained by some means, it should be possible to explore a wide variety of (strain induced) phenomena discussed in the domain of quantum field theory.\cite{Miransky, Zubkov_2} As we will discuss below, our STM topographs reveal that the crystals are indeed under strain and generate a pseudo-magnetic field the effect of which is also observed in our STM spectroscopy experiments.

\begin{figure}[h!]
		\includegraphics[width=0.5\textwidth]{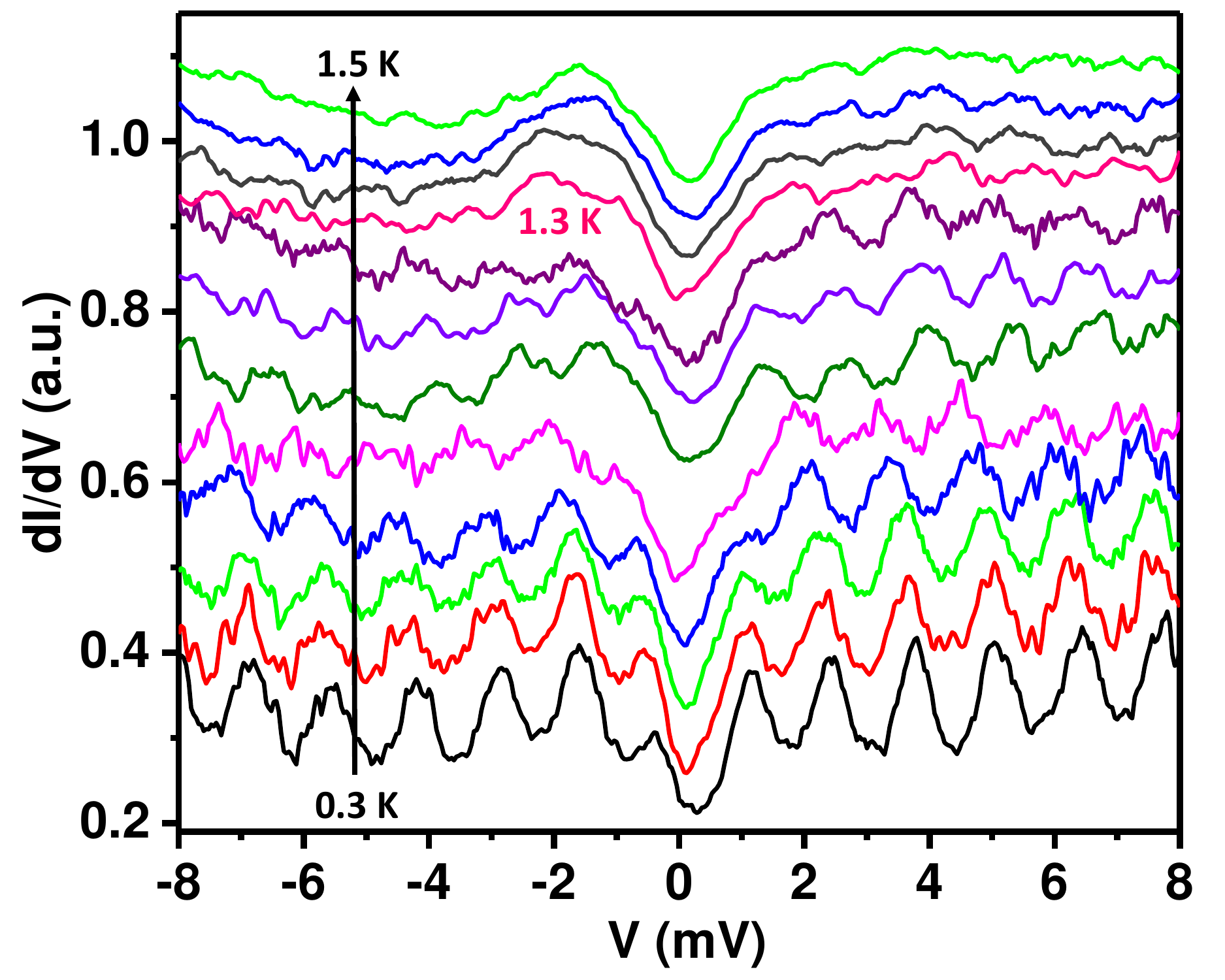}
		\caption{Temperature evolution of the STS data. The oscillations get vanished at 1.5 K.}	

\end{figure}

For the current studies, high quality single crystals of Re-doped MoTe$_2$ were grown by vapor transport technique. The scanning tunneling microscopy (STM) and scanning tunneling spectroscopy (STS) experiments were carried out in a low temperature, ultra-high vacuum (UHV) cryostat working down to 300 mK (Unisoku systems). First, a single crystal of Re-MoTe$_2$ was mounted in a low-temperature cleaving stage housed in the exchange chamber of the system where the crystal was cleaved by an $in-situ$ cleaver at 80 K in UHV ($10^{-11}$mbar). After cleaving, the crystal was immediately transferred by an UHV manipulator to the scanning stage which hangs from the end of the insert with three metal springs at low-temperature. The sample is biased and the tip is virtually grounded through the current pre-amplifier.

In Figure 1(a) we show a large area (700 nm x 700 nm) STM topographic image of one part of the surface captured at a temperature of 7 K. In this part, a large number of stripes are seen. The stripes occur periodically in real space with a modulation wavelength of approximately 35 nm. The bright-dark contrast of the stripes do not change when the sign of the tip-sample bias is reversed. The pattern does not change on varying temperature. On the other hand, a line profile (Figure 1(b)) drawn across the stripes show a periodic height variation of the order of $\sim 8 \AA$. These observations confirm that the stripes are not due to an electronic order, but due to the formation of atomic scale topographic modulations (ripples) on the surface of the single crystals. We have probed these ripples by changing scan angles, scan areas and other scan parameters to unambiguously confirm that they are real topographic features. The number of ripples ($n$) and their spacing ($\Delta l$) do not vary with changing the scan angle and for all scan angles, $n$ and $\Delta l$ change systematically with changing scan size. The ripples could have formed due to strain developed either during crystal growth or during the cleaving of the crystals prior to STM experiments. Such stripy surface ripples emerging from strain were earlier observed on single crystals of Bi$_2$Te$_3$. We will show later that such areas are maximally strained and the effect of a pseudo-magnetic field is most prominent in such areas. 

A closer inspection of the area (Figure 1(a)) also reveals the existence of another bunch of stripes formed along a different direction and with a much higher width. A corresponding line-cut is shown in Figure 1(c). In Figure 1(d), we show a different region away from that shown in Figure 1(a). In this region, the dense ripples have not formed but the stripes with larger width are present. These stripes might be due to formation of a Moire pattern induced by the strained surface of the crystal. Such strain-induced patterns were earlier seen in other two dimensional materials like graphite.\cite{HOPG} From the absence of the denser ripples in this area, it is trivial to conclude that this region is less strained than that shown in Figure 1(a). In fact, in this region, the strain responsible for the denser ripples (as in Figure 1(a)) are absent. We will show later that in these regions, the effect of the pseudo-magnetic field is not observed.  

When we zoom into a small area (20 nm x 20 nm) as shown in Figure 1(d), we observe the periodic arrangement of the atoms on the surface. Unlike in case of undoped MoTe$_2$, in Re-MoTe$_2$, our transport and STM measurements reveal the existence of a charge density wave (CDW) phase. The details of the CDW phase will be discussed elsewhere. At this length scale, the arrangement of the atoms on the surface is found to be identical at all regions under different levels of strain.

Now we focus on the local tunneling spectroscopy experiments. When we move the STM tip to points located in the low-strain area as shown in Figure 1(d) at a temperature of 310 mK, we obtain the spectra as shown in Figure 2(a). We show three representative spectra obtained at three points on the surface. All three of them show the evidence of a superconducting energy gap. Superconductivity in Re-MoTe$_2$ with 20\% Re doping is known to appear at $\sim$ 3.8 K. However, apart from the superconducting energy gap, no other prominent spectral features are observed. The details of the superconducting phase as probed by STM spectroscopic experiments will be discussed elsewhere. 

When we move the STM tip to different points on the ``high-strain" region represented by Figure 1(a), we obtain spectra as seen in Figure 2(b). At 310 mK, apart from the central dip due to superconductivity, distinct and strong peaks are seen in the $dI/dV$ vs. $V$ spectra. The three spectra shown in the figure are obtained from three different points in the same ``high-strain" region. The peaks are separated by an energy of $\sim$ 1.3 meV. These peaks do not appear in the spectra obtained from the ``low-strain" regions of the crystal. We have obtained spectra with such peaks on more than 50 points in different ``high-strain" regions. The position of the peaks remain almost at the same location with approximately 5\% variation which might be due to the variation of the strain itself from one ``high-strain" region to another. Since the $dI/dV$ spectrum at a given point is directly related to the local density of states (LDOS) at that point, the peaks in this case represent quantized energy levels. Such quantized levels appearing in presence of a large externally applied magnetic field have been observed in a number of topological systems in the past.\cite{Landau levels} These are the well-known Landau levels. In the present case, since the discretization takes place without any externally applied magnetic field, the peaks can be attributed to the Landau levels forming due to a strain-induced pseudo-magnetic field appearing in the system.  

Before attributing the discrete energy levels with the pseudo-magnetic field induced Landau levels, we have carefully considered other sources of such signal. For example, in STM experiments, oscillations may arise due to trapped impurity states. We have ruled out this possibility by performing measurements on ``high-strain" areas near and far from the impurities as seen in STM images. The oscillation pattern did not depend on that. In fact, in the ``high-strain" regions, we found the oscillations at all points that we measured. The role of other complex confinement effects is also ruled out because  in our data, the energy gap between two consecutive peaks remained nearly the same for a given spectra and almost same for all the spectra in a given "high-strain" region. This is not expected for an arbitrary confinement potential. Furthermore, Klein tunneling in Weyl and Dirac semimetals forbids the formation of such trapped/confined states.\cite{Klein tunneling} Role of superconductivity for the oscillations has also been ruled out as (a) the oscillation frequency and amplitude remained insensitive to an externally applied magnetic field and (b) in the ``low-strained" area superconductivity was found but no oscillations. In addition to the above-mentioned facts, the appearance of the oscillations in the ``high-strain" areas and the absence of the same in the ``low-strain" areas directly confirms that there is a direct correlation between the  observed oscillations and strain. 

\begin{figure}
\centering
\rotatebox[origin=c]{0}{\includegraphics[width=0.9\columnwidth]{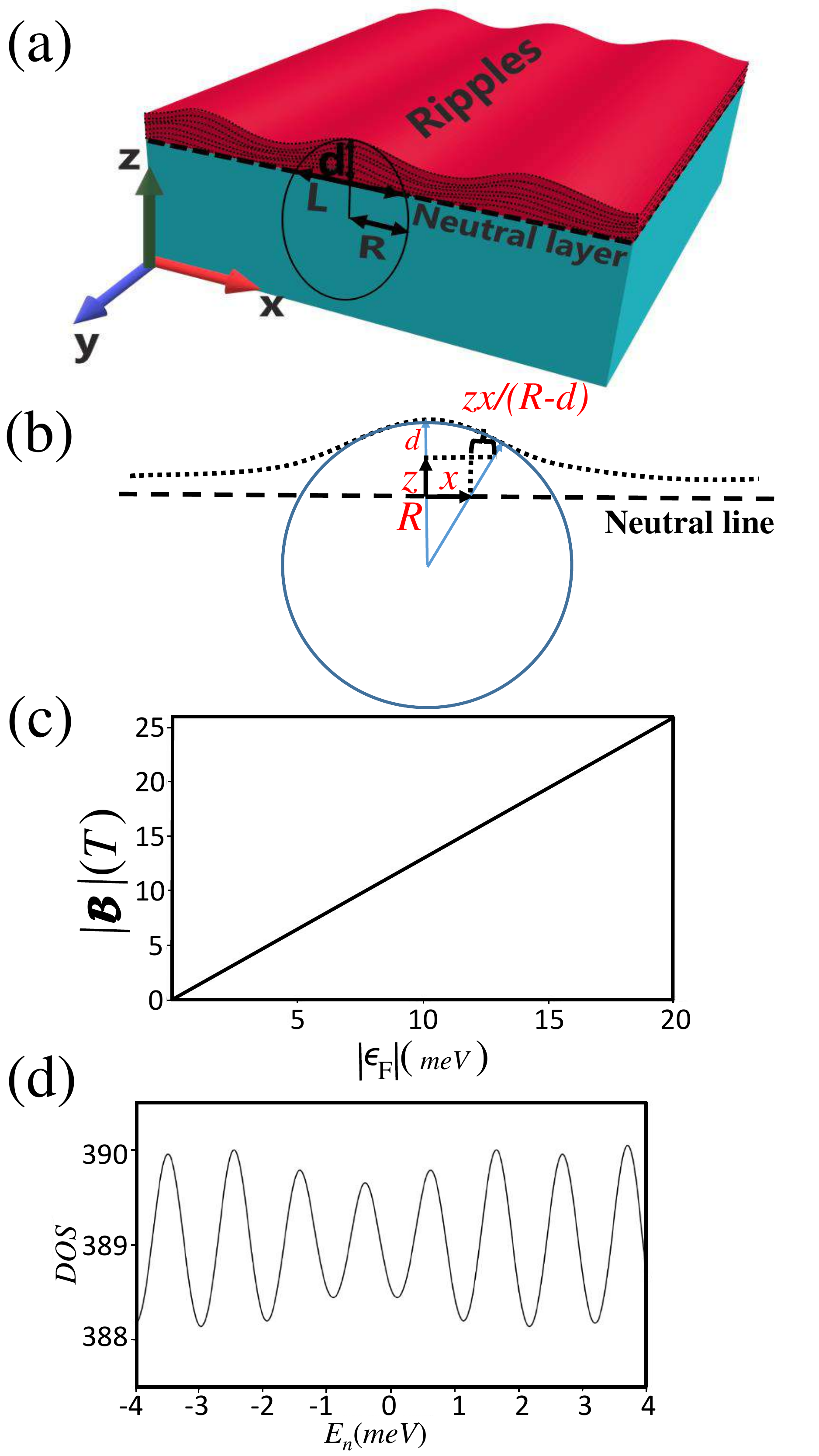}}
\caption{Theoretical model solution to the strained induced LLL spectrum. (a) A Schematic description of the setup, consisting of ripples with $d$-length sticking out from the neutral surface (first strain-less surface down the material). (b) A single ripple (black dotted line) and its tip with radius of curvature $R$. $x$ and $z$ are the displacement in the corresponding directions, giving a displacement field along $x$-direction as $u_x=zx/(R-d)$. (c) $\mathcal{B}$ vs $\epsilon_{\rm F}$ plot, from which order of pseudo-magnetic field is estimated semi-classically. (d) Density of states vs. Energy showing discrete Landau levels.
}
\label{thfig}
\end{figure}
Now it is imperative to theoretically understand the key observations namely, the Landau quantization of quasiparticle density of states emerging from a pseudo-magnetic field induced by strain in doped MoTe$_2$. We first provide a model framework of the observed strain (as manifested by ripples or periodic topographic modulation on the surface plane) in a type-II Weyl Hamiltonian, and follow it up with the study of corresponding LLs and oscillations in DOS in accordance with the experimental data. From the STM images (Figure 1), we find that the ripples are very much one-dimensional and periodic, with average ripple height of $8\r{A}$ (from the surface), and with about $L$=35 nm of inter-ripple distance.  Radius of curvature of the tip of ripple is $R$, and tip of $d\sim 41.3\r{A}$-height is sticking out from the neutral plane of the surface, and $L$ is the spatial extent of one ripple. We orient our coordinate system with $y$-axis lying along the ripple axis and $z$-axis is perpendicular to the neutral plane, as shown in Fig.~\ref{thfig}(a). From electronic structure point of view, the electron's hopping along the $y$-direction remains unchanged, while that along the $x$-axis is modulated in space, giving the essential gauge field in this case. 

We use a type-II Weyl Hamiltonian specific for MoTe$_2$\cite{DFTMoTe2,Rhodes} $H=d_0({\bf k})I_{2\times 2} + \sum_{\mu =1}^3 d_{\mu}({\bf k})\sigma_{\mu}$, where  $\sigma_{\mu}$ are the 2-component Pauli matrices, and $d_{\mu}$ are their anisotropic coefficients. The key difference between the type-I and type-II Weyl Hamiltonians arises from the onsite dispersion $d_0$ which is even and odd under inversion with respect to the Weyl points in the two cases, respectively. For a type-II system, this odd parity is responsible for tilting of the Weyl cone thereby giving rise to unique quantum properties. For convenience of illustration, we simplify the coefficients with a trivial rotation of the coordinate system from the lattice frame of reference to our above-mentioned set-up and obtain $d_0 = t_0 \sin(k_ya_y)-\epsilon_{\rm F}$, $d_{\mu} = t^{\prime}_{\mu} \sin(k_{\mu} a_{\mu})$, where $\mu=$1,2,3, and lattice constants $a_{\mu}$ are the lattice constants in the strained lattice, and $k_{\mu}$ are measured with respect to the Weyl points ${\bf Q}$ . $t_0$, and $t^{\prime}_{\mu}$ are the intra-species and inter-species hopping integrals, respectively. 

As mentioned before, the rippled structure induces a spatial dependence in the tight-binding hopping along the $x$-direction as  $t^{\prime}_{x}\sigma_{x}\rightarrow t^{\prime}_x(1-u_{xx})\sigma_{x}+\sum_{\mu^{\prime}\neq x}u_{\mu \mu^{\prime}}t^{\prime}_{\mu^{\prime}}\sigma_{\mu^{\prime}}$, where $u_{ij}$ is the stress tensor obtained from the displacement field vectors ${\bf u}$ as $u_{xx}=\partial u_x/\partial x$.\cite{Dirac,WeylQO} The displacement field vector can be approximated to the lowest order as ${\bf u}=[zx/(R-d),0,0]$, see Fig.~\ref{thfig}(b).To appreciate how the spatially modulated hopping renders a vector potential, it is convenient to use the long-wavelength limit of the Hamiltonian by substituting $\sin{(k_{\mu}a_{\mu})}\sim k_{\mu}a_{\mu}$, and expressing the tilt velocity as $\omega_{\perp} = t_{y}a_y$, and the Weyl fermion velocity as $v_{\mu}=t^{\prime}a_{\mu}$, we obtain the effective low-energy Hamiltonian as

\begin{equation}
H\approx (\omega_{\perp} k_y-\epsilon_{\rm F})I_{2\times 2} + \sum_{\mu=1}^3 v_{\mu}\sigma_{\mu}(k_{\mu}-e\mathcal{A}_{\mu}).
\label{WeylHam}
\end{equation}
(We set $\hbar=1$ for simplicity). The above equation resembles a typical low-energy Hamiltonian for massless fermions under a vector potential ${\bf \mathcal{A}}$, except here ${\bf \mathcal{A}}$ arises intrinsically from $t^{\prime}_{\mu}({\bf r})$. With a lengthy algebra, we estimate ${\mathcal{A}}=\frac{1}{e }\Big(\frac{z}{a_{x}(R-d)}{\rm tan}(Q_{x}a_x),0,\frac{x}{a_{z}(R-d)}{\rm tan}(Q_{z}a_z)\Big)$ where ${\bf Q} \sim 2\pi/a_x(0.18, 0.17, 0)$, is the location of the Weyl points in this system. With the experimental inputs of $3\%$ lattice strain giving $a_x=3.46\r{A}$,$a_z=13.86\r{A}$, $d=41.3\AA$ and $R=34d$ we obtain the magnetic field to be $\mathcal{B} \approx 3\it{T}$. The estimated magnetic field is large enough to separate the LL's above the intrinsic broadening scale of the system. However, the manifestation of the LL's into oscillations in DOS is required to overcome additional constrained posed by the peculiarities of the type-II Weyl fermions. In such Weyl fermion cases, we find that (a) the tilt velocity manifests into dispersive LLs and hence its bandwidth $W$ is required to be smaller than the LL spacing for its visualization, and 
(b) the tilt velocity must overcome a threshold value to commence closed Fermi pocket.

{\it (a) Dispersive Landau levels:} The LL's are split across the tilting term with its dispersive obtained from Eq. (1) as 
\begin{equation}
E_n^{\pm}(k) = \omega_{\perp} k_y-\epsilon_{\rm F} \pm \sqrt{(v_yk_y)^2  + 2(v_{\perp}/l_B)^2 n},
\label{LL}
\end{equation}
where the magnetic length $l_B=1/\sqrt{e\mathcal{B}}$, $v_{\perp}=\sqrt{\mid v_{x}v_{z}\mid}$. Interestingly, the LLs are chiral  with its velocity arising from the tilt velocity $\omega_{\perp}$. Such chiral cyclotron orbits arise with electric field perpendicular to the velocity\cite{Efieldgraphene}. Hence, the type-II Weyl semimetals can be understood to experience pseudo-magnetic and pseudo-electric field simultaneously under strain. The LL splitting can be appreciated in the limit of $2v_{\perp}/l_{B} > v_{y}k_y$ which is satisfied in the large magnetic field limit as achieved in our experiments.  The calculated DOS are shown in Fig.~\ref{thfig}(d) for $\mathcal{B}=3T$ and with all other band parameters set as in Ref.~\cite{DFTMoTe2,Rhodes}.

{\it (b) Closed FS formation:} According to the Lifshitz-Kosevich paradigm, the quantum oscillation requires the FS perpendicular to the field orientation to form closed contour, and the oscillation frequency is proportional to the FS area.\cite{Lifshitz} Type-II Weyl fermions however may fail to form closed FS pocket due to the tilt,\cite{TypeIIWeyeLL} unless the tilt velocity is substantially modified. The pseudo-magnetic field is aligned along the $y$-direction, and thus we focus on the condition for a closed FS in the $(k_x,k_z)$-plane. We find that the DFT-derived value of the band parameters for MoTe$_2$ fail to  give closed contour, However, the modifications of the hopping parameter $t'_{\mu}$ with the pseudo-magnetic field renders closed Fermi surface on the $(k_x,k_z)$-plane. Specific calculation is provided in the SI.  

In conclusion, we have provided direct experimental evidence of the emergence of a pseudo-magnetic field in a strained type-II Weyl semimetal Re-MoTe$_2$. We have observed clear oscillation of the density of states with energy due to the formation of Landau levels created by the strain-induced pseudo-magnetic field. In addition, we have theoretically elucidated the origin of the pseudo-magnetic field and the Landau level dispersion for the materials specific type II Weyl semimetal by building a strain-based model. The experimental observations are consistent with the theoretical calculations within this model. The experiments as well as the theoretical analysis reveal the strength of the strain-induced pseudo-magnetic field to be $\sim$ 3 Tesla.  

GS acknowledges the financial support from the Swarnajayanti fellowship awarded by Department of Science and Technology, Govt. of India. SK and AV acknowledges UGC for senior research fellowship (SRF). TD acknowledges financial support from the Infosys Science foundation under Young investigator Award.


\begin{thebibliography} {99}
	
			\bibitem{Weyl1} B. Yan, and C. Felser, \textit{Annu. Rev. Condens. Matter Phys.} \textbf{8}, 337-54 (2017).
	
			
			\bibitem{Hasan} M. Z. Hasan, S. -Y. Xu, and G. Bian,   \textit{Phys. Scripta} \textbf{164}, 014001 (2015).
			
			\bibitem{Weyl2} X. Wan, A. M. Turner, A. Vishwanath, and  S. Y. Savrasov, \textit{Phys. Rev. B} \textbf{83}, 205101 (2011).		

			\bibitem{Weyl3} A. A.Burkov, M. D. Hook,and  L. Balents, \textit{Phys. Rev. B} \textbf{84}, 235126 (2011).			
			
			
			\bibitem{Weyl4} H. Weyl, \textit{PNAS} \textbf{15}, 323-34 (1929).
			
		
			\bibitem{Weyl5} O. Vafek,and A. Vishwanath,\textit{Annu. Rev. Condens. Matt. Phys.} \textbf{5}, 83-112 (2014).			
			\bibitem{Weyl6}  W. W. -Krempa, G. Chen, Y. B. Kim, and L. Balents, \textit{Annu. Rev. Condens. Matt. Phys.} \textbf{5}, 57-82 (2014).			
			
			
			\bibitem{Weyl7} S. -Y. Xu et. al., \textit{Science} \textbf{349}, 613 (2015).
			

			\bibitem{Hasan_CA} S. Jia, S. -Y. Xu, and  M. Z. Hasan, Weyl semimetals, \textit{Nature Materials} \textbf{15}, 1140 (2016).			
			
			\bibitem{Cava_CA} J. Xiong, S. K. Kushwaha, T. Liang, J. W. Krizan, M. Hirschberger, W. Wang, R. J. Cava, and N. P. Ong, \textit{Science} \textbf{350}, 413-416 (2015).			
			
			\bibitem{Park_CA} P. Kim, J. H. Ryoo, and C. -H. Park, \textit{Phys. Rev. Lett.} \textbf{119}, 266401 (2017).			
			
		
		\bibitem{Pikulin} D. I. Pikulin, A. Chen, and M. Franz, \textit{Phys. Rev. X} \textbf{6}, 041021 (2016).
			
			
			\bibitem{Zubkov} M. A. Zubkov, \textit{Annals of Physics} \textbf{360}, 655-678 (2015).
			
			
			\bibitem{CKLiu} C. -X. Liu, P. Ye, and X. -L. Qi, \textit{Phys. Rev. B} \textbf{87}, 235306 (2013).	
			
			
			\bibitem{Cortijo_1} A. Cortijo, Y. Ferreir\'{o}s,  K. Landsteiner, and M. A. H. Vozmediano, \textit{Phys. Rev. Lett.} \textbf{115}, 177202 (2015).
			
			
			\bibitem{Cortijo_2} A. Cortijo, D. Kharzeev, K. Landsteiner, and M. A. H. Vozmediano, \textit{Phys. Rev. B} \textbf{94}, 241405(R) (2016).
			

			\bibitem{Franz} T. Liu, D. I. Pikulin, and M. Franz, \textit{Phys. Rev. B} \textbf{95}, 041201(R) (2017).			
			
			\bibitem{Arjona} V. Arjona, and M. A. H. Vozmediano, \textit{Phys. Rev. B} \textbf{97}, 201404(R) (2018).			
			
			\bibitem{Ilan} A. G. Grushin, J. W. F. Venderbos, A. Vishwanath, and R. Ilan, \textit{Phys. Rev. X} \textbf{6}, 041046 (2016).			
			

			\bibitem{Fujimoto} H. Sumiyoshi, and S. F. Fujimoto, \textit{Phys. Rev. Lett.} \textbf{116}, 166601 (2016).			
			
			
			\bibitem{TypeIIWeyeLL} S. Tchoumakov, M. Civelli, and M. O. Goerbig, \textit{Phys. Rev. Lett.} \textbf{117}, 086402 (2016).
					
			
			\bibitem{Graphene} F. Guinea, M. I. Katsnelson, and A. K. Geim, \textit{Nat. Phys.} \textbf{6} 30 (2010).
			

			\bibitem{Levy} N. Levy, S. A. Burke, K. L. Meaker, M. Panlasigui, A. Zettl, F. Guinea, A. H. C. Castro Neto, and M. F. Crommie, \textit{Science} \textbf{329}, 544 (2010).				
			
			
			\bibitem{Madhavan} Y. Okada, W. Zhou, D. Walkup, C. Dhital, S. D. Wilson, and V. Madhavan, \textit{Nature Communications} \textbf{3}, 1158 (2012)
			
			\bibitem{CFelser} Y. Sun, S. C. Wu, M. N. Ali,  C. Felser, and B. Yan, \textit{Phys. Rev. B} \textbf{92} 161107(R) (2015).
			
			\bibitem{Guguchia} Z. Guguchia et al., \textit{Nature Communications} \textbf{8} 1082 (2017).			
			

			\bibitem{Clarke} R. Clarke, E. Marseglia, and H. P. Hughes, \textit{Philosophical Magazine B} \textbf{38} 121-126 (1978).			
			
			\bibitem{XDai} Z. Wang, D. Gresch, A. A. Soluyanov, W. Xie, S. Kushwaha, and X. Dai, \textit{Phys. Rev. Lett.} \textbf{117}, 056805 (2016).
			

			\bibitem{Kaminski} L. Huang, T. M. McCormick, M. Ochi, Z., M. -to Suzuki Zhao, R. Arita, Y. Wu, D. Mou, H. Cao, J. Yan, N. Trivedi, and A. Kaminski, \textit{arXiv:1603.06482v1} (2016).			

			\bibitem{Qi} Y. Qi et al., \textit{Nature Communications} \textbf{7} 11038 (2016).			
			
			\bibitem{RPSingh} M. Mandal, S. Marik, K. P. Sajilesh, Arushi, D. Singh, J. Chakraborty, N. Ganguli, and R. P. Singh, \textit{Phys. Rev. Materials} \textbf{2} 094201 (2018).			
			
				
			\bibitem{Miransky} P. O. Sukhachov, E.V. Gorbar, I. A. Shovkovy, and V. A. Miransky, \textit{Ann. Phys. (Berlin)} \textbf{530}, 1800219 (2018).	
					
		
			\bibitem{Zubkov_2} A. Cortijoa, and M. A. Zubkov, \textit{Annals of Physics} \textbf{366}, 45-56 (2016).				
			
			\bibitem{HOPG} H. M. Guo, H. W. Liu, Y. L. Wang, H. J. Gao, H. X. Shang, Z. W. Liu, H. M. Xie, and F. I. Dai, \textit{Nanotechnology} \textbf{15}, 991 (2004).		
			

			\bibitem{Landau levels} P. Cheng et al., \textit{Phys. Rev. Lett.} \textbf{105}, 076801 (2010).	
			
			

			\bibitem{Klein tunneling} M. Jung et al., \textit{Nano Lett.} \textbf{18} 1863 (2018).			
			
			\bibitem{DFTMoTe2} Z. Wang et al., \textit{Phys. Rev. Lett.} \textbf{117}, 056805 (2016).
						
			
			\bibitem{Rhodes} D. Rhodes et al., \textit{Phys. Rev. B}  \textbf{96}, 165134 (2017).			
			

			\bibitem{Dirac} H. Shapourian, T. L. Hughes, and  S. Ryu, \textit{Phys. Rev. B} \textbf{92}, 165131 (2015).			
			

			\bibitem{WeylQO} T. Liu, D. I. Pikulin, and M. Franz, \textit{Phys. Rev. B}  \textbf{95}, 041201(R) (2017).			
				

			\bibitem{Efieldgraphene} V. Lukose, R. Shankar, and G. Baskaran, \textit{Phys. Rev. Lett.} \textbf{98}, 116802 (2007).			
			
			\bibitem{Lifshitz} I. M. Lifshitz, and A. M. Kosevich, \textit{Sov. Phys. JETP} \textbf{2}, 636 (1956).
			
	

\end{thebibliography}
\end{document}


\title{Supplementary Material for Generation of pseudo-magnetic field and consequent Landau level oscillations in intrinsically strained Re-MoTe$_2$\\}

\author{Suman Kamboj$^1$, Partha Sarathi Rana$^2$, Anshu Sirohi$^1$, Aastha Vasdev$^1$, Manasi Mandal$^3$, Sourav Marik$^3$, Ravi Prakash Singh$^3$, Tanmoy Das$^2$ and Goutam Sheet$^1$}

\email{goutam@iisermohali.ac.in}

\affiliation{$^1$Department of Physical Sciences, Indian Institute of Science Education and Research(IISER) Mohali, Sector 81, S. A. S. Nagar, Manauli, PO: 140306, India.}

\affiliation{$^2$Department of Physics, Indian Institute of Science, Bangalore 560012, India}

\affiliation{$^3$Indian Institute of Science Education and Research Bhopal, Bhopal, 462066, India}

\maketitle

High quality Re doped MoTe$_2$ crystals were cleaved by an \textit{in-situ} cleaver at 80 K under ultra-high vacuum (10^{-11} mbar).\\ 
\textbf{Additional STM images:}

\begin{figure}[h!]
\centering
\includegraphics[width=0.75\textwidth]{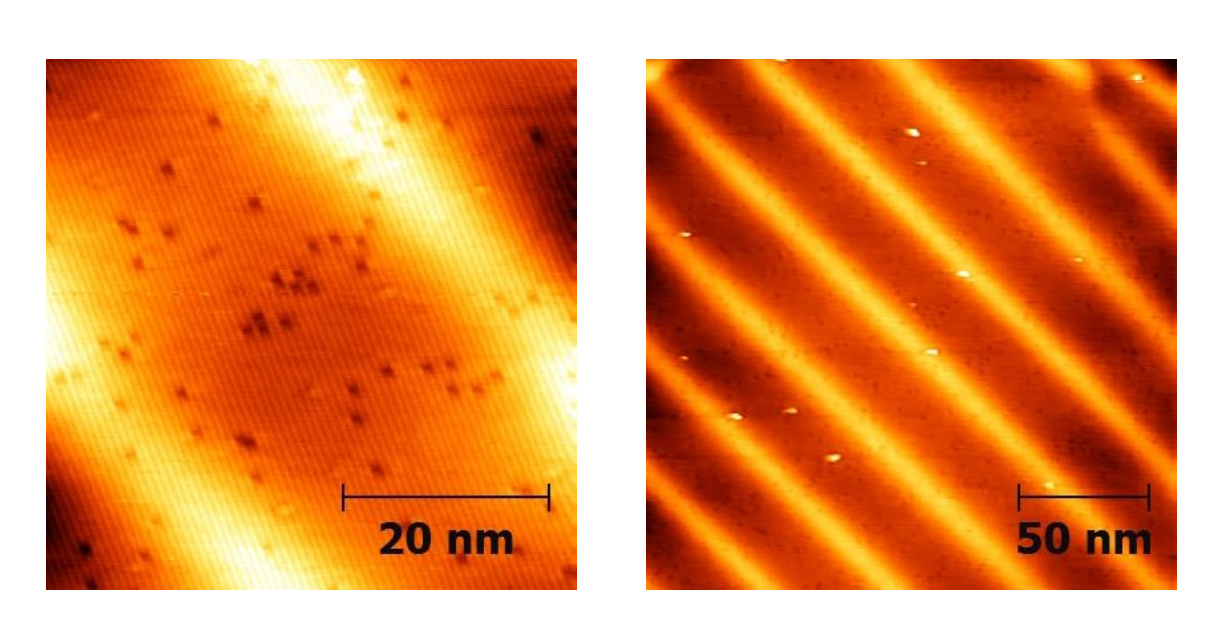}
\caption{60 nm x 60 nm and 200 nm x 200 nm STM topograph of the cleaved surface of Re-MoTe$_2$ obtained in a ``high-strain" region respectively. The periodic modulation due to strain-induced ripples is observed.}	
\end{figure}\\

\textbf{Additional STM spectra:}

\begin{figure}[h!]
\centering
\	\includegraphics[width=0.85\textwidth]{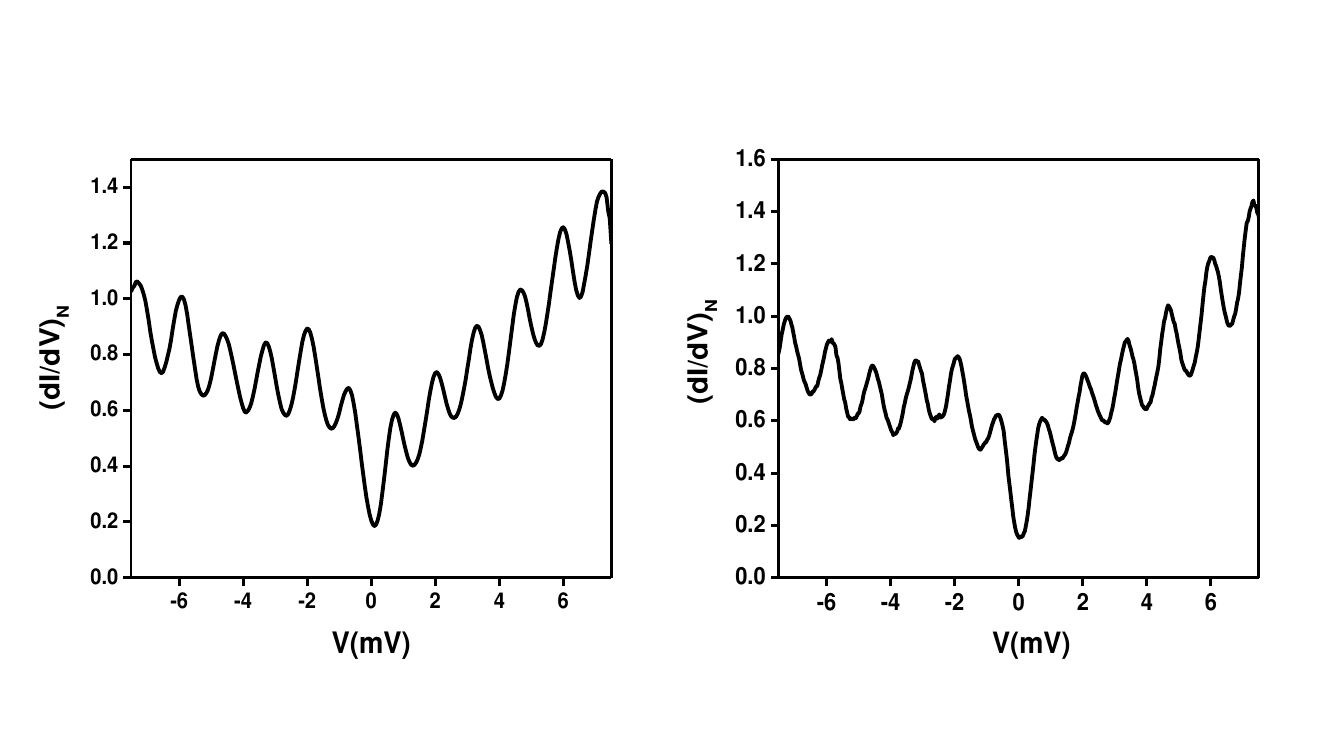}
\caption{ Differential conductance spectrum recorded at
310 mK showing LL oscillations along with a low-bias conductance
dip at two additional points in the high strained region.}	

\end{figure}\\

\textbf{Theoretical study of strain induced Landau levels in Re doped MoTe$_2$:}
We start with a specific model for type-II Weyl semimetal phase for MoTe$_2$ as presented in Ref.~\cite{DFTMoTe2,Rhodes} as
\begin{equation}
H({\bf k})=(\omega_{1}k_{x}+\omega_{2}k_{y}-\epsilon_{\rm F})I_{2\times 2}+(ak_{x}+bk_{y})\sigma_y+(ck_{x}+dk_{y})\sigma_{z}+ek_{z}\sigma_{x},
\label{eq2}
\end{equation}
In MoTe$_2$ system there are four Weyl nodes ${\bf Q}$ at $(0.1819,0.1721,0),\\(0.1300,0.0793,\pm 0.298),(0.1011,0.0503,0)$
where we use the same notation as in Ref.~\cite{DFTMoTe2} for the expansion parameters and a chemical potential ($\epsilon_{\rm F}$) is added to include the Re doping effect. We employ first a unitary transformation $R_y(\theta)=e^{i\theta/2 \sigma_y}$ with $\theta=3\pi/2$ on the Hamiltonian to obtain 
\begin{equation}
H({\bf k})=(\omega_{1}k_{x}+\omega_{2}k_{y}-\epsilon_{\rm F})I_{2\times 2}+(ak_{x}+bk_{y})\sigma_y+(ck_{x}+dk_{y})\sigma_{x}-ek_{z}\sigma_{z}.
\label{eq3}
\end{equation}
Next we give a rotation of the coordinate space with respect to the $k_z$-axis by an angle $\phi={\rm tan}^{-1}(\omega_2/\omega_1)$ to yield
\begin{equation}
H({\bf k})= (\omega_{\perp}k_y-\epsilon_{\rm F})I_{2\times 2}+d_y\sigma_y+d_x\sigma_{x} + d_z\sigma_{z}.
\label{eq4}
\end{equation}

where $\omega_{\perp}=\sqrt{\omega_1^2 + \omega_2^2}$, and $d_y=(a\cos{\phi} + b\sin{\phi})k_x + (b\cos{\phi} - a\sin{\phi})k_y$, and $d_x=(c\cos{\phi} + d\sin{\phi})k_x + (d\cos{\phi} - c\sin{\phi})k_y$, and $d_z=-ek_z$. With a further rescaling of the parameters ($a$, $b$, $c$, $d$, and $\phi$), we obtain our simplified Hamiltonian in the main text to be $d_{\mu} = v_{\mu}k_{\mu}$,  where $v_{x}=(b\cos{\phi} - a\sin{\phi})$, $v_y=(c\cos{\phi} + d\sin{\phi})$, and $v_z=-e$.The four Weyl nodes at (in $RLU$) $(0.0085,0.0498,0.101)$, $(0.0290,0.1696,0.1819)$, $(-0.2804,0.1284,0.1300)$,\\ $(-0.3071,0.0279,0.1300)$ in our rotated frame of reference. 

\subsection{Modeling strain induced magnetic field}

For a strained material, we define displacement field vector which determines the change in the lattice position (relative to original position): ${\bf u}=({\bf r^{'}}-{\bf r})$. According to the model shown in Fig.~, the lowest order term in the displacement of atoms from its equilibrium position is
\begin{align}
{\bf u}=\Big(\frac{zx}{R-d},0,0\Big), 
\label{eq16}
\end{align}
where $d$ is the depth of the neutral axis from the tip of the bent (along z axis). The strain is along the x-axis which gives the relation as 
\begin{align}
\alpha=\frac{a_x-a_{x0}}{a_{x0}}=\frac{d}{R-d}=\frac{3}{100}, 
\end{align}
where $\alpha$ is strain amount. Experimental value of strain is $\sim$3\%.

Under this displacement the hopping parameter changes as $t^{\prime}_{x}\sigma_{x}\rightarrow t^{\prime}_x(1-u_{xx})\sigma_{x}+\sum_{\mu^{\prime}\neq x}}u_{\mu \mu^{\prime}}t^{\prime}_{\mu^{\prime}}\sigma_{\mu^{\prime}}$, where $u_{xx}=\frac{\partial{u_{x}}}{\partial{x}}$. Now expanding the Hamiltonian around an Weyl node $\vec{Q}=\Big(0.0085(\frac{2\pi}{a_x}),0.0498(\frac{2\pi}{a_y}),0.1011(\frac{2\pi}{a_z})\Big)$, we get the vector potential :
\begin{align}
{\mathcal{A}}=\frac{1}{e }\Big(\frac{z}{a_{x}(R-d)}{\rm tan}(Q_{x}a_x),0,\frac{x}{a_{z}(R-d)}{\rm tan}(Q_{z}a_z)\Big),
\end{align}
pseudo magnetic field as 
\begin{align*}
{\bf \mathcal{B}}=\frac{1}{e }\Big(\frac{{\rm tan}(Q_{x}a_x) }{a_x(R-d)}-\frac{{\rm tan}(Q_{z}a_z) }{a_z(R-d)}\Big) \hat{y}, 
\end{align*}
where $a_x=3.45\r{A}$, $a_x=13.86\r{A}$, $R=1418\r{A}$, $d=\frac{R}{34}$. 
For $R=1418\r{A}$, $\mathcal{B} \approx 3\it{T}$. It is clear that from the last equation we can find the order of the pseudo-magnetic field by estimating the wavelength of ripples, ($35 nm$), depth of the neutral axis (d), and consequently the radius of curvature ($R(1418\r{A})$).

\subsection{Density of states}
If we choose a gauge, which leaves only $k_{y}$ and to be a good QM numbers ($\mathcal{B}$ is in y direction). The energy spectrum is,
\begin{align}
E_n^{\pm}(k) = \omega_{\perp} k_y-\epsilon_{\rm F} \pm \sqrt{(v_yk_y)^2 + 2(v_{\perp}/l_B)^2 n}
\end{align}
where $v_{\perp}=\sqrt{\mid v_{x}v_{z}\mid}$, 
Green's function:
\begin{align}
\mathcal{G(\omega)}&=\sum_{k_{y},n,\pm}\frac{1}{\omega-E_{n}^{\pm}(k_{y})+i\delta}\\
DOS&=-\frac{Im(\mathcal{G(\omega)})}{\pi}
\end{align}
with $k_{y}$ cutoff=$\pi/20$, and $n$ cutoff=200,we get 
peaks,which are $\sim 1meV$  apart, in $DOS$.

\subsection{Lifshitz-Kosevich results and condition for closed Fermi surface}
Fitting parameters in $eV$ $\AA$ units $\omega_{1}$=-3.39, $\omega_{2}$=0.58, $a$=0, $b$=0.78, $c$=2.6, $d$=-0.58, $e$-0.0045. The energy eigenvalues are
\begin{equation}
H({\bf k})= (\omega_{\perp}k_y-\epsilon_{\rm F})I_{2\times 2}+d_y\sigma_y+d_x\sigma_{x} + d_z\sigma_{z}.
\label{eq4}
\end{equation}
where $\omega_{\perp}=\sqrt{\omega_1^2 + \omega_2^2}$, and  $d_{\mu} = v_{\mu}k_{\mu}$,$v_{x}=(b\cos{\phi} - a\sin{\phi})$, $v_y=(c\cos{\phi} + d\sin{\phi})$, and $v_z=-e$.  

Let us consider a constant energy surface at an energy $\epsilon$, for the $E^{-}_{\bf k}$ state which gives the condition
\begin{align}
(\epsilon+\epsilon_{\rm F}-\omega_{\perp}k_{y})^2=v_{x}^2k_{x}^2+v_{y}^2k_{y}^2+v_{z}^2k_z^2, 
\label{constantE}
\end{align}
This equation can be cast into general equation of quadric surface (by completing square and shifting) as
\begin{equation}
\frac{k_{x}^2}{A}+\frac{k_{y}^2}{B}+\frac{k_{z}^{2}}{C}=1.
\label{eq3}
\end{equation}
where $A=\frac{G}{v_{x}^2}$, $B=\frac{G}{v_{y}^2}$, $C=\frac{G}{v_{x}^2}$,                 $G=(\epsilon+\epsilon_{\rm F})^2\Big[1+\frac{\omega_{\perp}^2}{v_{y}^2-\omega_{\perp}^2}\Big]$

The above equation gives a constant energy surface, which is  basically quadric surfaces. The contour we need is the cut of the quadric surface and the plane perpendicular to pseudo-magnetic field. This contour may not be closed depending on the nature of the quadric surface. The nature of quadric surface and therefore the contour can be determined by the signs of the coefficients in \eqref{eq3}. For most general case, we consider the closed contour of the electron to be an ellipse. After determining the coefficients, we come to the conclusion that to get an elliptic contour (closed), the orientation of magnetic field is along the $y$-axis. And the contour equation is,
\begin{align}
\frac{k_{x}^{2}}{\kappa_x^2}+\frac{k_{z}^{2}}{\kappa_z^2}=1,
\end{align}  
For most general case, we consider the closed contour of the electron to be an ellipse with its semi major, and semi minor axes being $\kappa_x$, and $\kappa_z$. Now, cyclotron frequency of an closed orbit(semi classical),
\begin{align}
\omega_c=\frac{2\pi e\mathcal{B}}{\hbar^2}\Big(\frac{\partial \mathcal{S}}{\partial\epsilon}\Big)^{-1},
\label{eq14}
\end{align}
where $\mathcal{S}(\epsilon)$ = area of the closed orbit of the constant energy contour $\epsilon$ and $e$=charge of electron, $\mathcal{B}$= magnetic field. So we have  
\begin{align*}
\mathcal{S}(\epsilon)&=\pi \kappa_x\kappa_z,
\end{align*}
Semi-classical quantization is applicable if, $\frac{\hbar\omega_c}{\epsilon_{\rm F}}\sim10^{-4}$, and 
\begin{align*}
\hbar\omega_c=\frac{2\pi e\mathcal{B}}{\hbar}\Big(\renewcommand{\bibname}{References}
\frac{\partial \mathcal{S}}{\partial\epsilon}\Big)^{-1}_{\epsilon_{\rm F}}.
\end{align*}
Now we have 
\begin{align*}
\mathcal{B}=\frac{\hbar^2\omega_c}{2\pi e}\Big(\frac{\partial \mathcal{S}}{\partial\epsilon}\Big)_{\epsilon_{\rm F}}.
\end{align*}
We put $\hbar\omega_c= 1.3$ meV, in close agreement with the experimental data and $\epsilon=0$ and in the above equation.